\documentclass[conference]{IEEEtran}

\usepackage[british]{babel}
\usepackage[utf8x]{inputenc}
\usepackage{graphicx, subfig, color}
\usepackage{algorithm, algorithmicx, algpseudocode}
\usepackage{amsmath, amsthm, amssymb, mathtools, amsfonts}
\usepackage{blkarray}
\usepackage{soul}
\usepackage{scrextend}
\usepackage{multirow}
\usepackage{cite}
\usepackage{booktabs}
\usepackage{textcomp}
\usepackage{xcolor}
\def\BibTeX{{\rm B\kern-.05em{\sc i\kern-.025em b}\kern-.08em
    T\kern-.1667em\lower.7ex\hbox{E}\kern-.125emX}}

\PassOptionsToPackage{hyphens}{url}\usepackage{hyperref}

\newcommand{\etal}{\emph{et al.}}

\usepackage{tikz}
\usepackage{textcomp}
\usepackage{hyperref}
\usepackage{lipsum}

\newcommand\copyrighttext{%
  \footnotesize \textcopyright  2022 IEEE.  Personal use of this material is permitted.  Permission from IEEE must be obtained for all other uses, in any current or future media, including reprinting/republishing this material for advertising or promotional purposes, creating new collective works, for resale or redistribution to servers or lists, or reuse of any copyrighted component of this work in other works. The official version can be found at
  \href{https://doi.org/10.1109/CSR54599.2022.9850307}{https://doi.org/10.1109/CSR54599.2022.9850307}}
\newcommand\copyrightnotice{%
\begin{tikzpicture}[remember picture,overlay]
\node[anchor=south,yshift=10pt] at (current page.south) {\fbox{\parbox{\dimexpr\textwidth-\fboxsep-\fboxrule\relax}{\copyrighttext}}};
\end{tikzpicture}%
}

\title{Ensemble of Random and Isolation Forests for Graph-Based Intrusion Detection in Containers }

\author{\IEEEauthorblockN{Alfonso Iacovazzi and Shahid Raza }
\IEEEauthorblockA{\textit{RISE Research Institutes of Sweden}\\
Stockholm, Sweden \\
\{name.surname\}@ri.se}
}

\IEEEoverridecommandlockouts
\begin{document}

\maketitle
\copyrightnotice

\begin{abstract}
We propose a novel solution combining supervised and unsupervised machine learning models for intrusion detection at kernel level in cloud containers. In particular, the proposed solution is built over an ensemble of random and isolation forests trained on sequences of system calls that are collected at the hosting machine's kernel level. The sequence of system calls are translated into a weighted and directed graph to obtain a compact description of the container behavior, which is given as input to the ensemble model.
We executed a set of experiments in a controlled environment in order to test our solution against the two most common threats that have been identified in cloud containers, and our results show that we can achieve high detection rates and low false positives in the tested attacks.
\end{abstract}


\section{Introduction}\label{sec:intro}

Virtualization based on containers has gained a lot of interest during last few years. Containers have become popular since they offer a lightweight system abstraction which includes sufficient components for running stand-alone and isolated cloud applications.

Despite the attention paid to security issues in the cloud computing environment, containers have been demonstrated vulnerable to several attacks \cite{mattetti2015securing,abed2015intrusion}. In general scenarios, a container is considered healthy at the start of its operation, and later it is infected with a malware. But in some cases, the container images which are stored in repositories and made available to the community (e.g., dockerhub), might be already infected with a malware. The malware can then be activated silently when the container image is installed and executed. Cryptohijaking malware is one example of malicious software that could be pre-injected in a container image. When executed, the cryptominer hiddenly shares the computation resources with the legitimate application. Additional examples of malicious code that can run inside containers are: Backdoors, Ransomware, External file transfer malware, etc.

According to some recent reports \cite{Backdoored}, several images available on dockerhub which have been downloaded millions of times, have been found with a malicious code pre-installed. Other reports showed that many servers in telecommunications, media, healthcare, hosting containers have been found infected with cryptocurrency miners. Cryptohijaking attacks are the widest threat targeting cloud containers nowadays.

During application deployment and execution in the containers, a malware process can hiddenly start and run in the background, this may lead to resource exhaustion, privacy breaching, and service disruption. Cloud service providers need to detect malicious activities within containers to protect its resources (storage, bandwidth, etc), protect user resources (data, web services, etc), and avoid the spread of the attack from one container to the others. At the same time, service providers cannot implement intrusive anomaly detection strategies which look at the information inside the container's resources.

Intrusion detection is commonly known as a hard and long-standing problem. It becomes more challenging in public cloud computing services when the monitoring system has not direct access to the resources assigned to  customers and their activities within the container processes. In addition, accurate monitoring cloud services is challenging because of the complex architectures and the large-scale of resources to monitor. For this reason, advanced detection solutions can play a crucial role in detecting malicious behaviors by analyzing the data related to the containers and observable at the hosting operating system which share the kernel with the hosted containers. Some features that a machine learning-based detector may exploit for this purpose can be derived by isolating and processing container-based system calls. System calls are the communication instrument between the running application and the Linux kernel of the hosting machine and can be captured by the cloud provider without accessing the container resources directly.
In this paper we provide a mechanism for (i) classifying container workload behaviors and (ii) detecting malicious behaviors within the containers. Our solution, working at the hosting operating system, is based on a mix of supervised and unsupervised machine learning models. While the (supervised) random forest algorithm is able the delineate the differences among the multiple container workloads which may run on the hosting machine, the (unsupervised) isolation forest model will identify anomalous behaviors in the containers. We exploit the anonymous walks embedding algorithm to extract a compact representation of the sequences of system calls which is given as input into the intrusion detection model. The use of graph representation by means of anonymous walks embedding allows to incorporate the hidden relations and dependencies between system calls within a time window.
To the best of our knowledge, we are the first to use anonymous walks for embedding a sequence of system calls into a machine learning model in the context of intrusion detection.

The contributions of this research work are as follow:
\begin{itemize}
  \item We provide a description of a data repository for intrusion detection in the context of cloud containers, which has been collected in a emulated environment running selected Docker workloads;
  \item We introduce a graph representation of the sequence of system calls collected at the hosting machine's kernel level which is processed by the random walks and anonymous walks algorithms to extract features of the container's behaviors;
  \item We present the design of a practical intrusion detection mechanism which incorporates the capability of distinguish among multiple container workloads. Our solution works on three stages and is based on an anonymous walks data embedding stage followed by a random forest classification module and an ensemble of independent isolation forests;
  \item Finally, we provide some results obtained by testing our solution on two datasets: one resulting from a public repository and another obtained from traces collected for three types of workloads.
\end{itemize}

The rest of the paper is organized as follows; We provide the state of the art related to this domain in Section \ref{sec:relworks}. Section \ref{sec:background} provides background information on cloud containers, graph theory, random and anonymous walks algorithms, random forest and isolation forest models, which are required to understand the solution proposed in this work. We describe our graph-based intrusion detection system in Section \ref{sec:model}. The setup for data collection in a cloud environment is described in Section \ref{sec:setup}. Section \ref{sec:results} presents the obtained results, and finally our conclusion is provided in Section \ref{sec:concl}.

\section{Related works}\label{sec:relworks}

Intrusion detection systems (IDSs) have been extensively studied during last decades, and several surveys on the achievements in this area have been produced \cite{liao2013intrusion,borkar2017survey,hindy2018taxonomy}. 
Some of the surveys available in the state-of-the-art aim their attention at a specific application domain: e.g., some authors review the IDS solutions for Internet of Things \cite{sherasiya2016survey,chaabouni2019network}, Can \etal~ focus on IDSs in wireless sensor networks \cite{can2015survey},  Sultana \etal~ analyse the research works in SDN-based networks \cite{sultana2019survey}. 

IDSs are usually classified into two categories: (i) host-based intrusion detection systems (HIDSs) which monitor and analyse the devices from their internal behavior \cite{liu2018host} and (ii) network-based intrusion detection systems (NIDSs) which monitor and analyse the network traffic to detect any malicious behavior \cite{chaabouni2019network}. In this work, we focus on HIDSs.

In Unix-like operating systems, system calls are the primary source of data used as input for detecting malicious behaviors in HIDSs \cite{liu2018host}. Monitoring the system calls to detect unusual behaviors in computer systems has been done since the nineties \cite{hofmeyr1998intrusion}.

Considering the large amount of calls that are usually generated in a system, multiple algorithms were implemented to extract information carried by the list of system calls. Some common methods are: the n-gram formatting \cite{khreich2017anomaly}, sliding window algorithm \cite{eskin2001modeling}, bag-of-words model \cite{kang2005learning}, and term frequency-inverse document frequency (TF-IDF) method \cite{chen2005application}. All these strategies get a sequence of  system calls as input and output a compressed representation of it.

One of the representation methods that has gained a lot of interest during last decade is the ``Bag of System Calls'' (BoSC) which was introduced for the first time by Kang \etal~ in 2005 \cite{kang2005learning}. Since then, BoSC was widely adopted as an effective and efficient way to represent list of system calls in intrusion and anomaly detection \cite{chen2005application}.

With the recent advancements of cloud systems which enable the execution of virtual applications over rented cloud resources, intrusion detection strategies based on the analysis of system calls at the hosting machine level has gained a lot of interest \cite{ates2018taxonomist,karn2020cryptomining}. 
Thus, a cloud service provider can build a scalable and efficient mechanism to monitor in parallel multiple containers running on the same machine with a single instance of intrusion detection. However, results provided so far show that only few solutions obtain acceptable detection rates.
Ates \etal{} \cite{ates2018taxonomist} developed the ``Taxonomist,'' a technique based on traditional classification models (e.g., random forest, decision trees, and SVM) which is able to distinguish the behavior of different application running on virtualized systems. The proposed mechanism receives as input the information coming from the resource utilization of each application. The proposed classification model is able to recognize unknown applications with high accuracy; however, they did not show what would happen if a malicious code is executed in parallel to the benchmarked applications.
Karn \etal{} \cite{karn2020cryptomining} proposed a supervised approach based on ensemble learning and decision trees for extracting cryptomining signature for explainability. They obtain high accuracy in detecting the cryptoming software behavior, however the solution is based on a supervised model with a focus on cryptomining and it may fail to detect new and unknown malicious behavior.

Finally, a framework for proactive intrusion detection was introduced by Gupta \etal{} \cite{gupta2017supervised}. The proposed model is based on a two stage LSTM-based algorithm aiming at analyzing how the resource utilization varies over the time and classifying normal and abnormal behavior in cloud containers. However, this solution yet obtains high false positive rates (about 10\%), which is undesirable in systems that are monitoring an high number of processes.

Unsupervised approaches have shown interesting results in detecting unknown malicious behavior in several domains \cite{casas2012unsupervised,zhang2006anomaly,zoppi2021unsupervised}. Cui and Umphress \cite{cui2020towards} recently proposed an unsupervised model for detecting attacks in containerized applications. Their solution is based on LSTM autoencoders to detect malicious behaviors. However, their results show high False positive and True positive rates which suggest that much more effort should be put in place to identify unsupervised solutions that would work in cloud environments.

\section{Background}\label{sec:background}

\subsection{Cloud Containers}
A container is an abstraction of executable software that includes whatever is necessary to run an application, such as source code, tools, data, dependencies, etc. Any specific application, workload, or task can be executed in a self-contained container, isolated from the rest of the hosting system. The lightweight nature of containers enables system efficiency and scalability since multiple containers can be easily configured, executed, customized, and monitored for different applications and purposes on the same hosting machine.

The market in this domain is pretty prosperous and multiple vendors are competing to provide the best solutions in term of cloud container infrastructures: Kubernetes engine by Google, Elastic Container Service by Amazon, Microsoft Azure, and Docker are the main container management tools available in the market \cite{bernstein2014containers}.

Docker is top player in this sector \cite{merkel2014docker}; in fact, more than 7 million developers use its tools for developing their applications. Some of main features that contribute to the widespread use of Docker are: (i) Docker is an open source containerization platform that offers integrated and automated security policy, (ii) it enables Continuous Integration/Continuous Deployment (CI/CD), (iii) it offers a public library including plenty of Docker images, (iv) it is simple to use, etc. For all these reasons, we chose Docker as the container infrastructure used in our experiments.

\subsection{System calls}
A system call is the procedure in which a program running on the operating system requests a service from the kernel. System calls are classified by type and an ID number is assigned to each type (from 0 to 322). Read, write, open, close are some of popular system calls in Unix, Unix-like and other POSIX-compliant operating systems. The mapping function between system call types and their ID numbers can be found on the searchable Linux syscall table \cite{syscall-table}.

\subsection{Graphs, Random and Anonymous Walks}\label{subsec:anonwalks}
We use the syntax $G=(V,E,X)$ to represent a weighted directed graph, where $V=\{v_1,v_2,\dots,v_N\}$ denotes a set of $N$ nodes, $E\subseteq V\times V$ denotes the set of edges, and $X\in\mathbb{R}_{N\times N}$ denotes the matrix of edge weights.

Given a graph $G$, we can generate a random walk graph $R=(V,E,P)$ with the following weight assignment: for every edge $e=(v,w)\in E$, the weight $p_e$ is equal to $x_e/\sum_{\epsilon\in E_{v,out}}{x_\epsilon}$, where $x_e$ is the weight of edge $e$ in graph $G$ and $E_{v,out}$ is the set of edges outgoing from node $v$.

A random walk of length $l$ is an ordered sequence of nodes $u_1, u_2,\dots, u_l$, in which the step $e_i=(u_i,u_{i+1})$ in the walk is randomly selected with a probability $p_{e_i}$ computed as the ratio between the weight of the edge $e_i=(u_i,u_{i+1})$ in the graph $R$ and the sum of the weights of the outgoing edges from $u_i$. The selection of the next node in the sequence is independent from the set of neighbors of the last node in the sequence.

Anonymous walks (AWs) were defined by Ivanov and Burnaev \cite{ivanov2018anonymous} to provide an instrument for learning data-driven graph embeddings.

Given a random walk $w=(u_1, u_2,\dots, u_k)$ of length $k$, the positional function $\phi(w, u_i)=(p_1,p_2,\dots,p_l)$ is defined as the list of all positions $p_j\in \mathbb{N}$ of $u_i$ occurrences in the random walk $w$.

For the random walk $w$, a corresponding anonymous walk is defined as the sequence of integers $a = (\alpha(u_1),\alpha(u_2),\dots,\alpha(u_l))$ where $\alpha(\cdot)$ is the function $\alpha(u_i) = \min{\{p_j\}}_{p_j\phi(w, u_i)}$.

Given the set $A_l$ of all possible anonymous walks of length $l$, and let $p(a_i)$ be the probability that  the $i$-th anonymous walk in $A_l$ exists in $G$, the Anonymous walk embedding is defined as the vector of the list of probabilities of $A_l$.

We refer the readers to the manuscript by Ivanov and Burnaev  \cite{ivanov2018anonymous} for further details about the anonymous walks embedding.

\subsection{Random Forest Classifier}
The Random Forest (RF) Classifier is a classification algorithm based on the ensemble learning \cite{ho1998random}. The algorithm aims at creating a multitude of decision trees, where each of them is generated from a randomly selected subsets of the training data. A node in a tree represents an hyperplane that divides the hyperspace where the training data samples fall into; on the other side the end nodes correspond to a specific region of the hyperspace. The output from the decision trees in the forest are combined by means of a bagging meta-learning algorithm. We refer the readers to the manuscript by Ho \cite{ho1998random}  for further details.

\subsection{Isolation Forest}
The Isolation Forest (IF) algorithm was proposed by Liu \etal{} as an unsupervised model-based method for isolating anomaly instances in a dataset rather than profiling the regular behaviors of the normal data \cite{liu2008isolation}. The idea behind the model assumes that the anomalies in a dataset are usually rare and isolated from the rest of data. Starting from this intuition, Authors built an algorithm in which the set of instances in the dataset are recursively and randomly partitioned until instances are isolated. Usually, more outliers in a dataset require a smaller number of iteration in order to be isolated. The recursive partitioning process can be translated into a binary tree (iTree) in which random partitioning iterations are mapped into tree nodes, and the leaves of the tree correspond to the dataset instances. The leaves with the shortest path in the tree will probably correspond to the outliers.

In the IF algorithm, an ensemble of iTrees is built during the training phase by random and recursive partitioning. In the testing phase, an anomaly score is given to the tested instances defined as:
\begin{equation}\label{eq:itscore}
    s(x,n) = 2^{\frac{-E(h(x))}{c(n)}}
\end{equation}
where $x$ is the tested instance, $n$ is the number of instances in the dataset, $E(h(x))$ is the average path length in the tree computed over the ensemble of iTrees for $x$, and c(n) is the average path length in an iTree with $n$ instances. We refer the readers to the manuscript by Liu \etal{} \cite{liu2008isolation}  for further details about IF algorithm. 
\section{Intrusion detection in cloud containers}\label{sec:model}
\subsection{Problem statement}
We consider a scenario in which a cloud service provider offers resources as containerized services. In a given hosting server there might be running multiple containers belonging to different customers and sharing the kernel of the hosting operating system. A monitoring application running on the hosting machine analyzes the samples of sequences of system calls for each hosted container and the extracted features are given as input to the intrusion detection algorithm. The machine learning models were earlier trained with clear traces corresponding to the workloads running on the server. The monitoring system should be able to (i) recognize that a workload effectively running in a specific container corresponds to the workload for the expected assigned application, and (ii) identify any anomalous/malicious behavior within the container. If an anomaly or workload mismatch is detect, the monitoring system will raise an alert and eventually may trigger resource preserving actions.

\subsection{Model description}

Since features resulting from the BoSC do not take into account the dependencies among adjacent system calls in a sequence, we  exploit a graph-based representation which is able to incorporate those dependencies while preserving the information in the frequency of occurrences; in addition graph theory offers some mechanisms to obtain a compact feature set.

Our intrusion detection system works in three main stages. During the first stage, the system calls collected at the hosting machine are processed and transformed into a graph representation based on anonymous walks.

The data representation resulting from the first stage is embedded into an RF classifier which is trained with normal behavior classes only; each class includes instances collected for a specific containerized workload running in regular setting with no compromise. Let $N$ be the number of containerized applications in the training set, in the last stage of the algorithm, the output of the RF classifier is given to $N$ independent IF instances each of them separately trained with a dataset made of samples belonging to one normal behavior class and contaminated with anomalies randomly extracted from the remaining normal classes. We selected random and isolation forests in our system since they run efficiently on large datasets and offer high accuracy in many classification problems.

\begin{figure}
\centering
	\includegraphics[width=0.4\columnwidth, trim=0cm 0cm 0cm 0cm, clip]{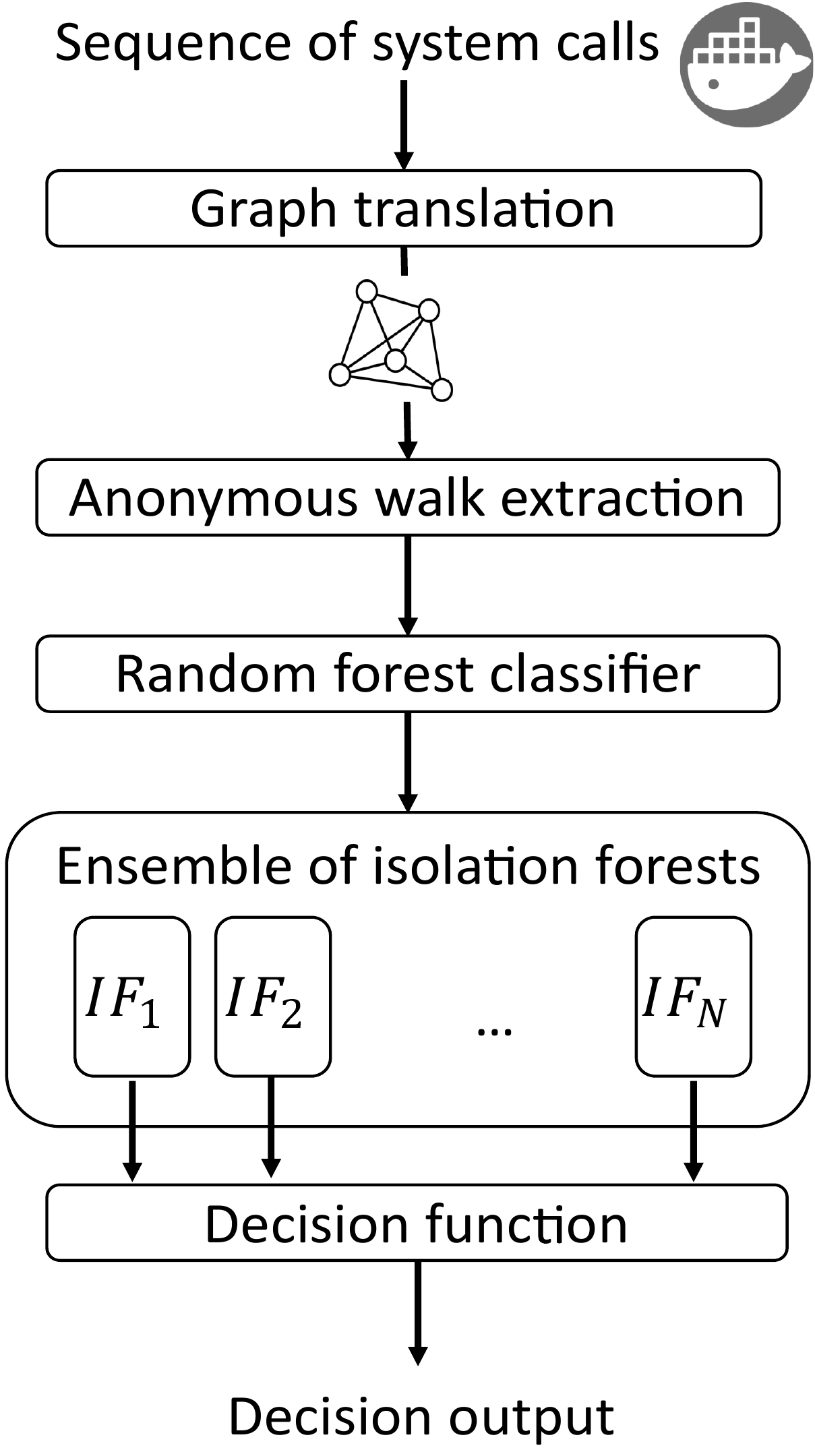}
	\caption{Block diagram of the proposed model.}
	\label{fig:algo}
\end{figure}

The three stages of our algorithm are depicted in Figure \ref{fig:algo}. Details of the three stages are provided in the next subsections.

\subsubsection{First Stage: Graph representation of System calls}
We introduce a representation of the sequence of system calls based on graph structures. Starting from a time period of duration $T$, we first extract the sequence $\mathcal{S}$ of system call's IDs observed during the period. Let $C_{IDs}$ be the set of the system call's IDs \cite{syscall-table}, we generate a graph $G_\mathcal{S}=(V_\mathcal{S},E_\mathcal{S},X_\mathcal{S})$ for the sequence of system calls where $V_\mathcal{S} = \{ v_i\}_{i\in C_{IDs}}$, $E_\mathcal{S}=\{(v_i,v_j)\}_{i,j\in V_\mathcal{S}}$, and $X_\mathcal{S}$ is the weight matrix where the $(i,j)$-th value is the weight for edge $(v_i,v_j)$ and is computed as number of times the bi-gram $(v_i,v_j)$ occurs in the sequence $\mathcal{S}$.
Given the graph $G_\mathcal{S}$, we compute its representation vector $W_\mathcal{S}=[w_1, w_2, \dots, w_L]$ based on the random walk embedding process introduced in Section \ref{subsec:anonwalks}.

\subsubsection{Second Stage: Random Forest Classifier}

The representation vector resulting from the first stage is given as input to the RF classifier. The classifier is trained against samples belonging to the $N$ normal classes. The training dataset is provided to the classifier with its labels. The model in the testing phase outputs two pieces of information: (i) a set of probabilities that the input sample belongs to each class in the training set, and (ii) the decision about the class for the input sample. The set of probabilities is passed to the next stage. The decision about the class is passed to the monitoring system, and might be used to trigger any decision about the container, if for example the class of the workload running in the container does not match the expected class.

\subsubsection{Third Stage: Isolation Forest}

The probabilities given by the second stage are imputed to an ensemble of $N$ IF modules for generating $N$ anomaly scores. Each IF is trained from a dataset made of samples processed by the first two stages and belonging to one normal class only and contaminated with samples resulting from the other normal classes. The output of the $n$-th IF module trained on the $n$-th class outputs an anomaly score for that specific class.

\subsubsection{Final Decision}
The $N$ anomaly scores from the ensembles of IFs are combined to make a decision of classification based on the following rules:
\begin{itemize}
  \item if all anomaly scores are below the threshold but the $n$-th one, the instance is classified as belonging to class $n$;
  \item if all anomaly scores are below the threshold, the instance is classified as anomaly;
  \item if more than one score are above the threshold, the instance is classified as anomaly.
\end{itemize}

\section{Experimental setup}\label{sec:setup}

We executed a set of experiments in a controlled environment in order to collect system call logs for the purpose of testing intrusion detection models. In particular, we setup three different Docker container workloads enabling different types of computational loads: (i) data analytics, (ii) media streaming, and (iii) web search application. We exploited the implementations (v3.0) of cloud applications provided in the benchmark suite for cloud services by Cloudsuite \cite{palit-demystifying-cloud-benchmarking}. The provided benchmarks are based on real-world software stacks and represent real-world setups.

Each workload was tested against three scenarios: (i) benign, (ii) malicious with Cryptomining software, (iii) malicious with backdoor and remote control.

For each scenario, multiple experiments were executed; the stream of system calls generated at the hosting machine level were registered for each executed experiment. 1000 experiments were executed for each scenario, for a total of 9000 instances.

Experiments were executed in isolated virtual machines running Ubuntu 20.04 LTS operating system. Docker Engine was used to containerize the selected applications starting from their Docker images \cite{merkel2014docker}.

\subsection{Workloads}
\subsubsection{Data analytics}
This workload generates a Hadoop cluster executing a set of data analysis tasks based on the MapReduce framework \cite{dean2008mapreduce}. The cluster includes a single master container running the driver program and multiple slave containers running the mappers and reducers. The tasks executed by the cluster includes: (i) filtering and transformation of the data, (ii)  data aggregation, and (iii) data classification based on Naive Bayes. Each experiment were executed by  selecting randomly the number of slaves (a value between 2 and 6).

\subsubsection{Media streaming}
This workload generates a Docker cluster including a streaming server and a streaming client. The server container is based on NGINX and hosts several videos of different quality and lengths. The client container, based on httperf, generates a mix of requests for different contents from the server.

\subsubsection{Web search}
The Web search workload provided by Cloudsuite relies on the Apache Solr search engine framework. It generates a server container running the Apache Solr index node and a client container simulating multiple real-world clients that send requests to the index node.

\subsection{Scenarios}
\subsubsection{Benign}
In this scenario, the workloads were executed according to the specifications provided by the developers. The docker images were pulled directly from the Cloudsuite repository and locally installed for the execution.

\subsubsection{Malicious with Cryptomining software}
The benchmark images for the three workloads were injected with malicious code running a cryptomining application. We focused on the Monero cryptomining tool, which has been identified as the main one injected for hidden mining in cloud containers since it uses privacy-enhancing algorithms to provide users' anonymity. The tool is provided in the Docker image, and the mining code is injected in the workload entrypoint so as it is activated as soon as an experiment starts. The mining is performed throughout the all duration of each experiment.

\subsubsection{Malicious with Backdoor and remote control}
The benchmark images for the three workloads were equipped with a userland rootkit (Evil rabbit \cite{rootkits}) including SO injection capabilities. The rootkit enables the hidden remote management of the container by a malicious third party. The rootkit is provided in the Docker image and activated at the starting of the experiment. A bash script, randomly drawing linux commands from a predefined list, run on a container (which was chosen among the containers running in a workload except the target one) and transmits the commands to the target container for remote control and file transfer.

\subsection{Experiment execution and data collection}
For each experiment, a VM was started, and all the required packages and Docker images were loaded on the VM. At the beginning of an experiment, a Docker network was instantiated, then all containers required in the workload were created and started. The containers were all connected to the same Docker network. The \emph{perf} tool, which is traditionally installed by default in Unix-like operating systems, was used to record system calls at the hosting VM level \cite{perf}. In particular, we monitored: the master container in the data analytics workload, the NGINX server in the media streaming workload, and the index node container in the web search workload. These containers are the ones that run the malicious code in the corresponding scenarios. Selective monitoring was possible by filtering the system calls based on the container's IDs.

\section{Performance analysis}\label{sec:results}
We tested our model on two different datasets: a dataset (Mix-2022) obtained by the traces collected in a controlled environment, as described in Section \ref{sec:setup}, and a dataset (Cui-2020) resulting from a public repository provided by Cui and Umphress \cite{cui2020towards}. The repository is made of traces captured in a Docker platform running a containerized MySQL server as sample application. Random commands were generate to interact with the server. The system call logs were collected under two scenarios: one with regular and clear application execution and the other with the application under attack. Developers implemented the following attacks: Brute Force Login, DockerEscape, Malicious Script, Meterpreter, Remote Shell, SQL Injection, and SQL Misbehavior. The repository contains 20 traces of system call logs for the benign container execution and 10 traces for each tested attack. We refer the readers to the manuscript by Cui and Umphress  \cite{cui2020towards} for further details about their experiment and data collection.

\subsection{Evaluation metrics}
Given a testing dataset consisting of $M_{n}$ samples from the normal behavior class and $M_{a}$  samples from a generic attack class, let $M_{TP}$ be the number of samples from the attack class that were correctly classified as anomalies (true positive), and $M_{FP}$ be the number of samples from the normal behavior class that were incorrectly classified as anomalies (false positive). In order to evaluate the performance of our algorithm, we used the following traditional metrics: (i) the true positive rate $TPR = \frac{M_{TP}}{M_a}$, (ii) the false positive rate $FPR = \frac{M_{FP}}{M_n}$, (iii) the precision $PR = \frac{M_{TP}}{M_{TP}+M_{FP}}$, and (iv) the score $F1=\frac{2\cdot PR\cdot TPR}{PR + TPR}$.

\subsection{Baseline}
We analyze the performance of the proposed solution by compering its results against a traditional models based on (i) Support Vector Machines (SVM) for the multi-class classification in the first stage and (ii) Local Outlier Factor and One-class SVM for second stage outlier detection, receiving as input the bag of system calls computed over a predefined observation period.

\subsubsection{Bag of System Calls}
A ``bag of system calls'' (BoSC) is a representation of the list of system calls observed by the kernel during a fixed period of time. Let  $S=\{s_i\}_i=1,\dots,N$ be the set of the $N$ types of different system calls for the operating system, and $L=(l_1,\dots,l_{K})$ be the sequence of $K$ system calls observed during the period of time $T=[t_{start},t_{end})$, with $l_k\in S$ for $k_1,\dots,K$. The bag of system calls $B_T$ for the time period $T$ is defined as $B_T=(b_1,b_2,\dots,b_N)$ where $b_i$ is the number of occurrences of the system call $s_j$ in the sequence $L$. BoSC-based representations are widely used in intrusion detection problems, as highlighted in Section \ref{sec:relworks}.

\subsubsection{Support Vector Machines}
Support Vector Machine (SVM) is a state-of-the-art supervised machine learning model commonly used as baseline in the intrusion detection domain, since it provides good and stable performance in classification and anomaly detection problems.

In an SVM classifier, samples in a training dataset are represented as point in an n-dimensional space. The SVM training algorithm aims at identifying the hyperplanes that partition the hyperspace in subspaces and separate the classes in the hyperspace. The hyperplanes are defined starting from the support vectors, i.e. data points in the training set that are closer to an hyperplane. By adding or removing support vectors, the algorithm influences the position and orientation of the hyperplanes so as to maximize the distance between the support vectors and the hyperplanes (margin). The hyperplanes resulting from the training algorithm will be the decision boundaries to determine the class of a given data sample.

In One-class SVM outlier detection, the algorithm aims at defining the hyperplanes that separate all the samples in the training set from the origin of the hyperspace.

We refer the readers to the original manuscripts by Boser  \cite{boser1992training} and Scholkopf \cite{scholkopf1999support} for further details about SVM-based classification and outlier detection.

\subsubsection{Local Outlier Factor}

The Local Outlier Factor (LOF) algorithm is an unsupervised anomaly detection method proposed by Breuning \etal \cite{breunig2000lof}. The algorithm computes for each data point an outlier score that measures the relative density deviation between the data point and its local neighbors; when the local density of the data point is lower than the density of its nearest neighbors, the data point is considered an outlier. 
We refer the readers to the original manuscripts by Breuning \etal \cite{breunig2000lof} for further details.

\subsection{Dataset creation and processing}
Starting from the original system call logs recorded and locally  stored, we randomly selected 10 not overlapping slots of duration $T=10$ sec for each trace in our repository so we could generate dataset ``Mix-2022'' made of 10 thousand samples for each class. Instead, for the traces available in Cui-2020, we randomly selected up to 15 not overlapping  slots of duration $T=10$ sec for each trace, obtaining a dataset made of 733 samples. 

Both datasets were split into training and testing datasets. In dataset Mix-2022, the samples belonging to the ``benign'' classes were split according to the following proportions: 70\% into the training and 30\% into the testing dataset; all the attack samples were allocated into the testing dataset. We derived four different training sets: S-set containing 70\% samples of the training set to train the (supervised) classifier, and the U(k)-sets (for $k=1,2,3$)  made of (i) all the samples in the training set belonging to the class $k$ and 2.5\% of the remaining samples in the training set belonging to the other classes (contamination). The U(k)-sets are used to train the unsupervised anomaly detection algorithm.

For dataset Cui-2020, we adopted the following process: the samples belonging to the ``benign'' class were split according to the 70-30 proportion, and all the attack samples were allocated into the testing dataset, as for dataset Mix-2022. The training dataset was contaminated with samples randomly generated and made of features drawn by independent probability mass functions obtained by the feature values in the training set. We set a contamination rate equal to 5\%.

\begin{figure*}\centering
    \subfloat[EoF - Dataset ``Mix-2022.'']{\includegraphics[width=0.48\columnwidth, trim=0cm 0cm 0cm 0cm, clip]{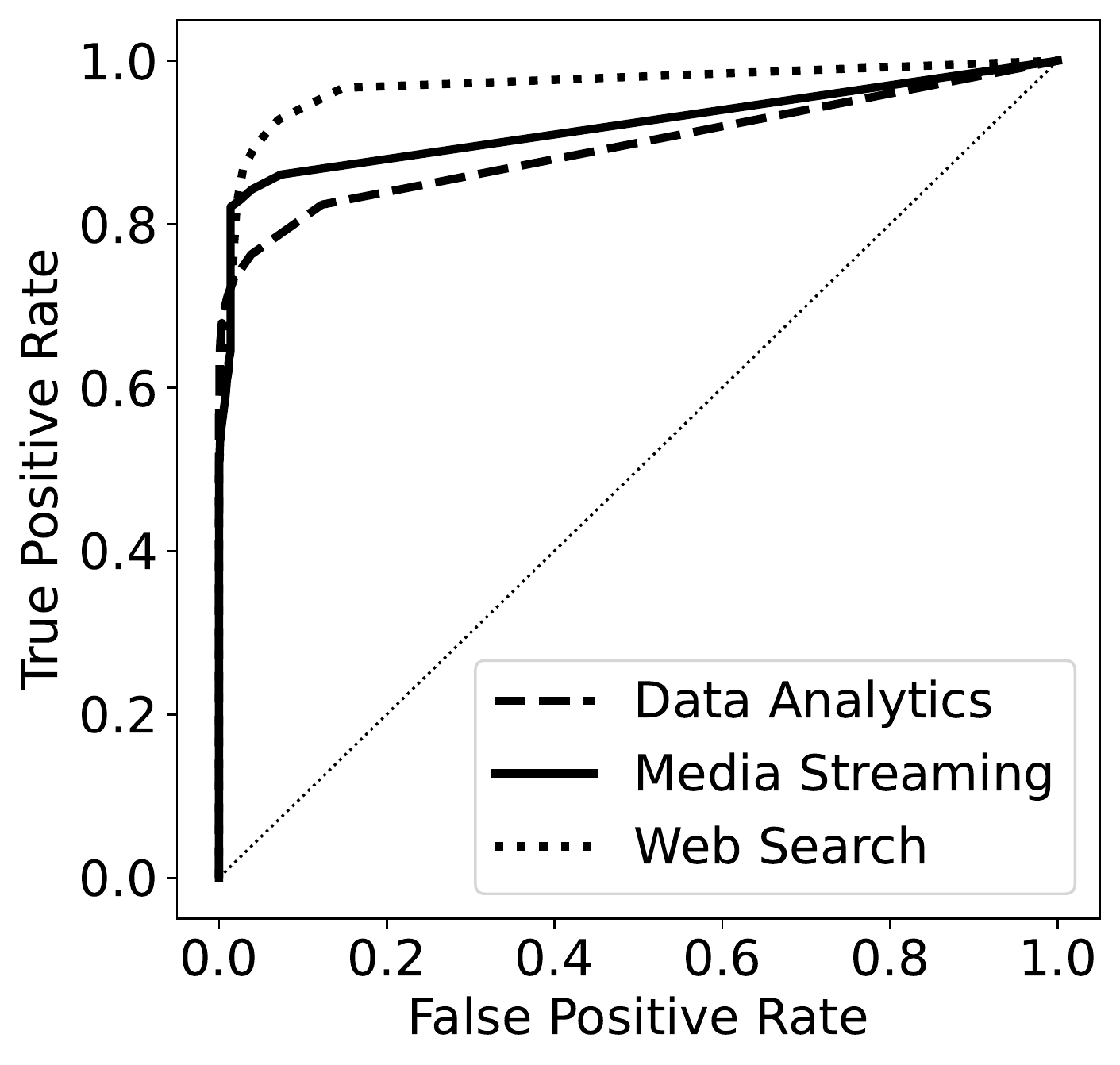}\label{fig:rocgid}}\quad
    \subfloat[SVM - Dataset ``Mix-2022.'']{\includegraphics[width=0.48\columnwidth, trim=0cm 0cm 0cm 0cm, clip]{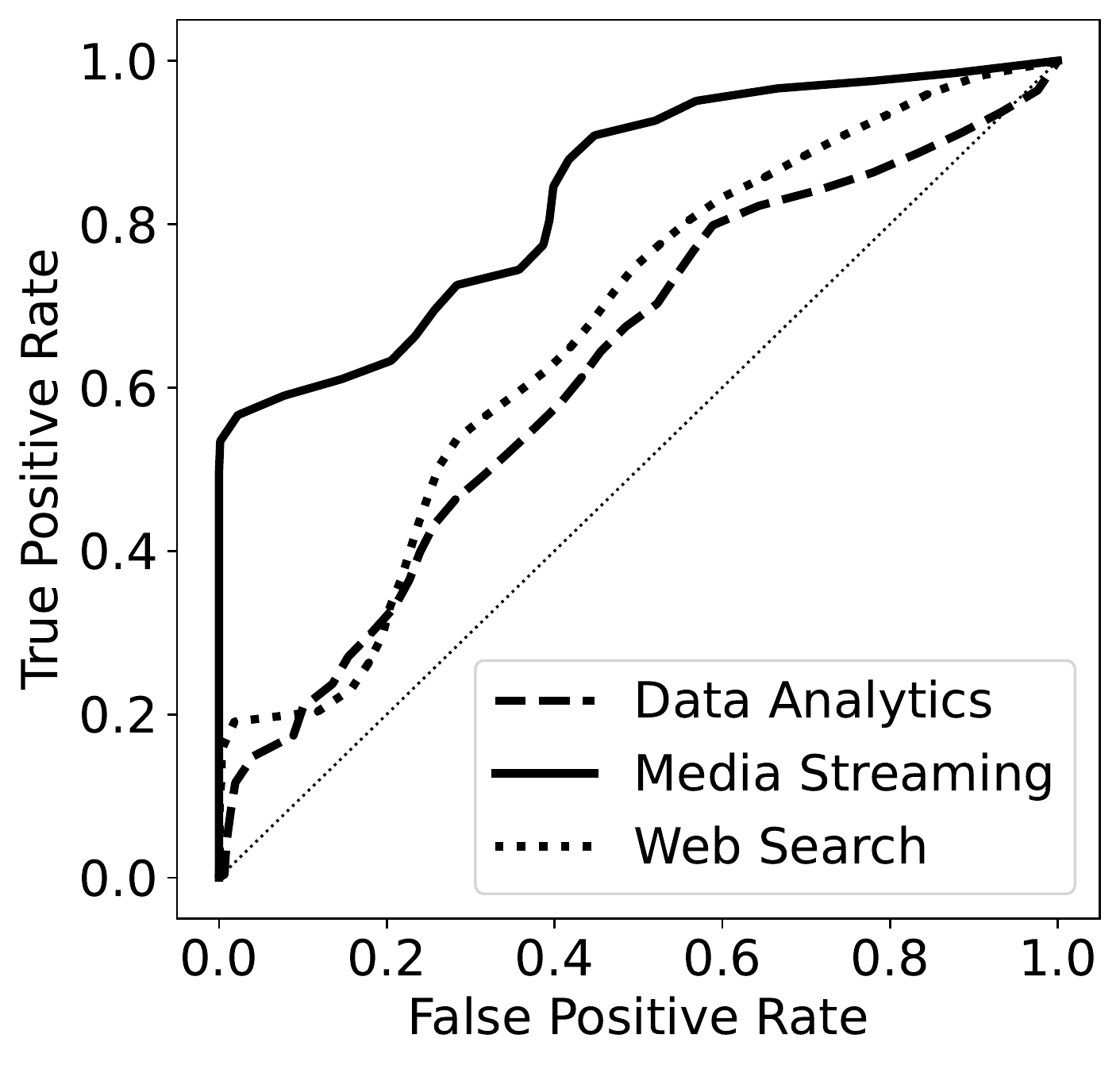}\label{fig:rocbasesvm}}\quad
    \subfloat[LOF - Dataset ``Mix-2022.'']{\includegraphics[width=0.48\columnwidth, trim=0cm 0cm 0cm 0cm, clip]{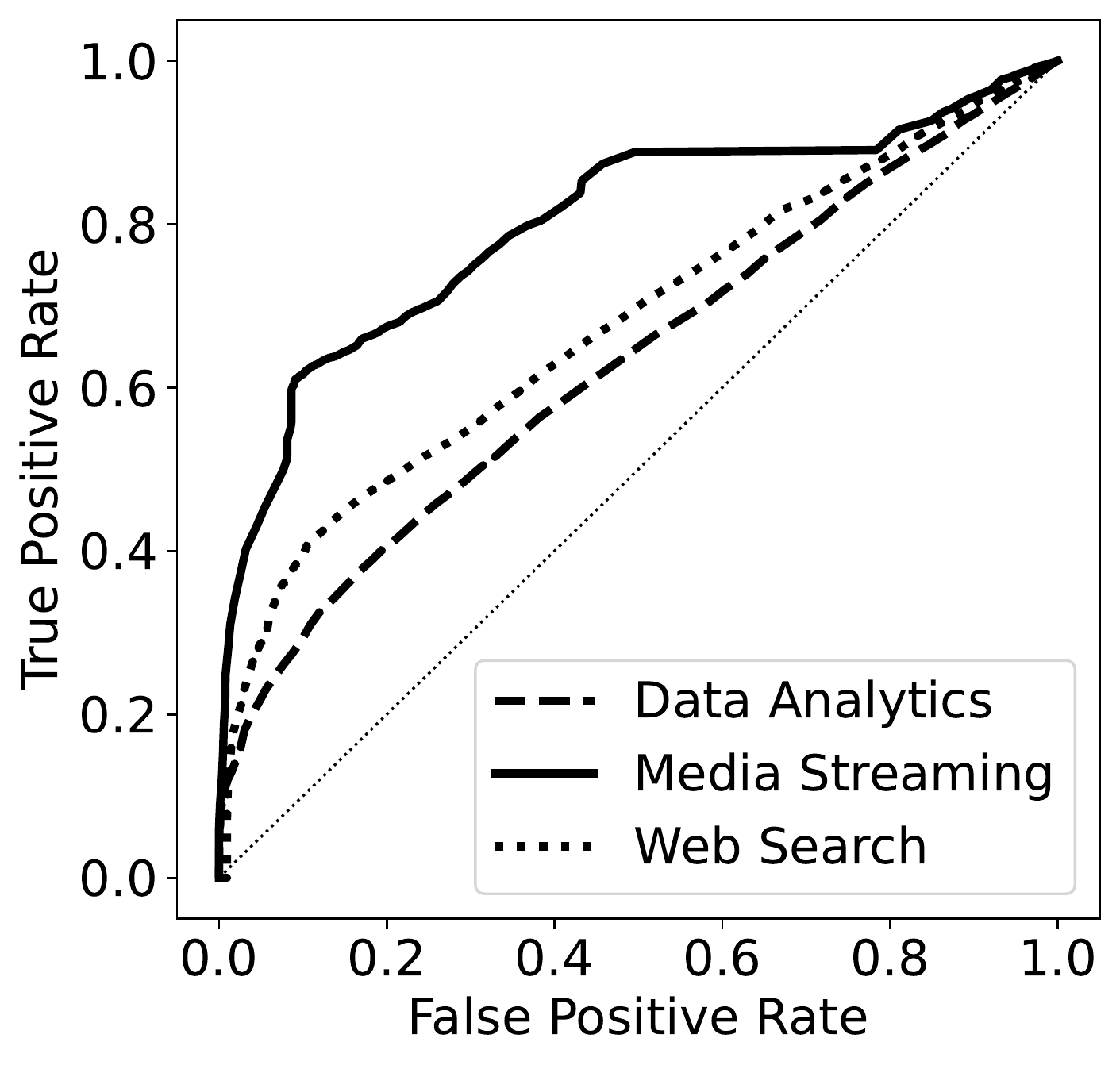}\label{fig:rocbaselof}}\quad
    \subfloat[Dataset ``Cui-2020.'']{\includegraphics[width=0.48\columnwidth, trim=0cm 0cm 0cm 0cm, clip]{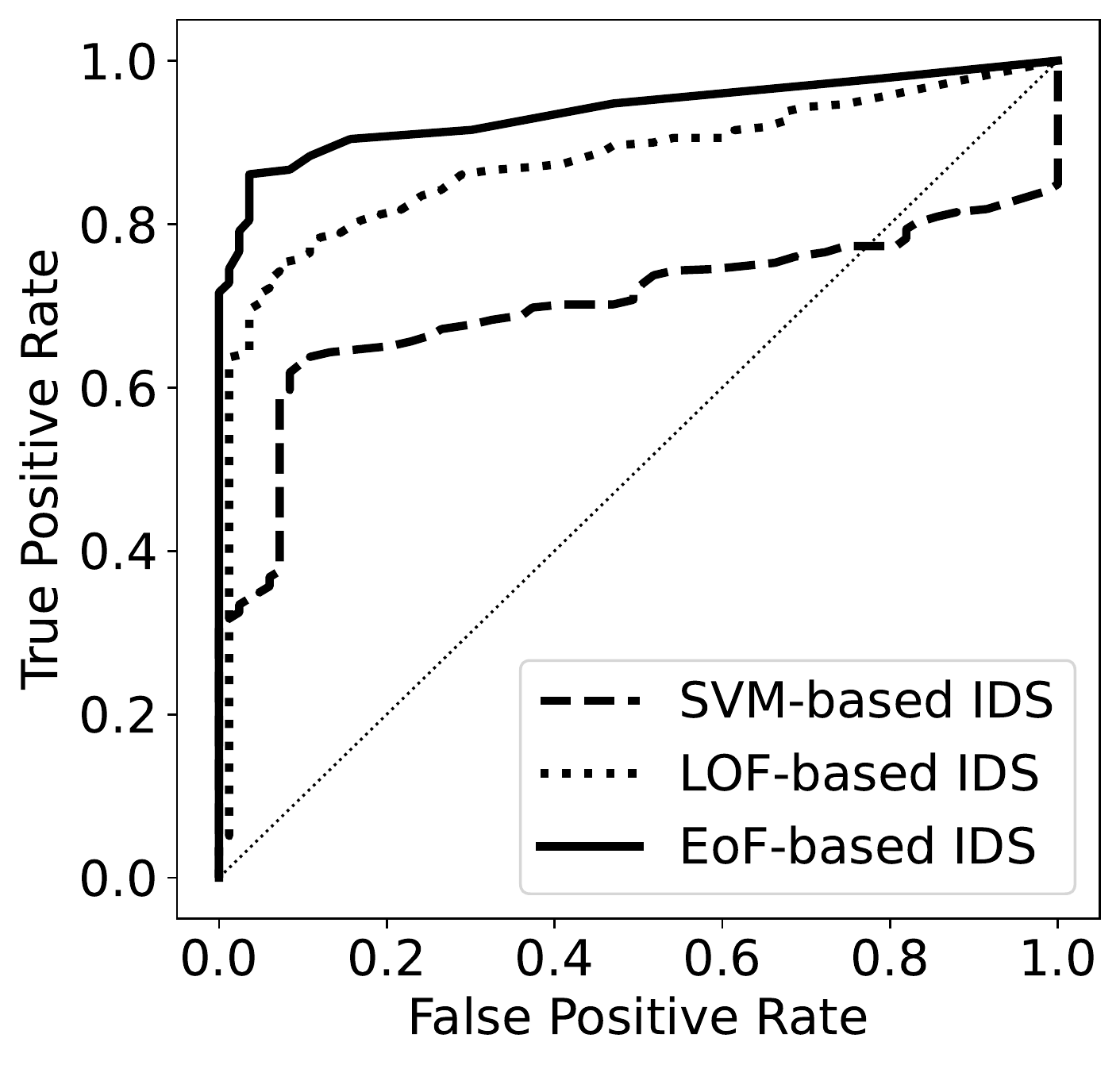}\label{fig:roccui}}
    \caption{ROC function for the three models.}\label{fig:roc2022}

\end{figure*}

\begin{table*}
\setlength{\tabcolsep}{6pt}
    \centering
    \caption{Performance comparison of the three models for dataset ``Mix-2022.''}\label{tab:tfprfirst}
    \begin{tabular}{ lccccccccc}
        \toprule
                &  \multicolumn{3}{c}{Ensemble of forests} & \multicolumn{3}{c}{One-class SVM} & \multicolumn{3}{c}{LOF}\\
        \midrule
                &  TPR & PR & F1 &  TPR & PR& F1 & TPR & PR& F1 \\
        \midrule
Data Analytics - C-mining	&	0.881	&	0.975	&	0.926	&	0.235	&	0.910	&	0.374	&	0.150	&	0.960	&	0.260	\\
Media Streaming - C-mining	&	1.000	&	0.982	&	0.991	&	1.000	&	0.982	&	0.991	&	0.996	&	0.818	&	0.054	\\
Web Search - C-mining	&	0.999	&	0.966	&	0.982	&	0.369	&	0.982	&	0.528	&	0.276	&	0.999	&	0.998	\\
Data Analytics - Backdoor	&	0.661	&	0.967	&	0.785	&	0.085	&	0.786	&	0.154	&	0.028	&	0.989	&	0.106	\\
Media Streaming - Backdoor	&	0.678	&	0.974	&	0.800	&	0.159	&	0.897	&	0.270	&	0.056	&	0.970	&	0.430	\\
Web Search - Backdoor	&	0.857	&	0.960	&	0.906	&	0.004	&	0.118	&	0.008	&	0.004	&	0.309	&	0.008	\\
        \bottomrule
    \end{tabular}
\end{table*}

\begin{table*}
\setlength{\tabcolsep}{6pt}
    \centering
    \caption{Performance comparison of the three models for dataset ``Cui-2020.''}\label{tab:tfprsecond}
    \begin{tabular}{ lccccccccc}

        \toprule
                &  \multicolumn{3}{c}{Ensemble of forests} & \multicolumn{3}{c}{One-class SVM} & \multicolumn{3}{c}{LOF}\\
        \midrule
                &  TPR & PR & F1 &  TPR & PR& F1&  TPR & PR& F1 \\
        \midrule
Brute Force Login	&	0.493	&	0.949	&	0.649	&	0.093	&	0.438	&	0.154	&	0.067	&	0.357	&	0.112	\\
DockerEscape	&	0.769	&	0.938	&	0.845	&	0.410	&	0.640	&	0.500	&	0.590	&	0.719	&	0.648	\\
MaliciousScript	&	0.711	&	0.964	&	0.818	&	1.000	&	0.894	&	0.944	&	1.000	&	0.894	&	0.944	\\
Meterpreter	&	1.000	&	0.958	&	0.979	&	1.000	&	0.836	&	0.911	&	1.000	&	0.836	&	0.911	\\
RemoteShell	&	0.973	&	0.973	&	0.973	&	0.959	&	0.886	&	0.921	&	0.959	&	0.886	&	0.921	\\
SQLInjection	&	0.610	&	0.959	&	0.746	&	0.961	&	0.892	&	0.925	&	0.857	&	0.880	&	0.868	\\
SQLMisbehavior	&	0.863	&	0.969	&	0.913	&	0.014	&	0.100	&	0.024	&	1.000	&	0.890	&	0.942	\\
        \bottomrule
    \end{tabular}
\end{table*}

\begin{table}
\setlength{\tabcolsep}{6pt}
    \centering
    \caption{Comparison of the FPRs.}\label{tab:fprates}
    \begin{tabular}{ lccc}
        \toprule
                &  EoF & SVM & LOF\\
        \midrule
Data Analytics - Benign	&	0.045	&	0.046	&	0.012	\\												
Media Streaming - Benign	&	0.036	&	0.037	&	0.001	\\												
Web Search  - Benign	&	0.071	&	0.061	&	0.017	\\												
\midrule																			
Cui-2020	&	0.024	&	0.108	&	0.108	\\

        \bottomrule
    \end{tabular}
\end{table}

\subsection{Results}
We implemented the models by using sklearn-learn library v0.24.2 in Python v3.5.2. Original Python implementation of anonymous walk embeddings algorithm \cite{ivanov2018anonymous} was adopted to translate the graph representation of system calls into feature vectors. We used anonymous walks of length 4 which produced vectors of 15 features. Both random and isolation forests were initialized with 100 estimators each. The SVM and LOF models were initialized with default configuration.

Tables \ref{tab:tfprfirst} and \ref{tab:tfprsecond} present the results obtained respectively for datasets Mix-2022 and Cui-2020. The tables list the TPR, PR, and F1 computed for each attack separately, while the FPRs are shown in Table \ref{tab:fprates}. The FPRs are computed for each workload in dataset Mix-2022, separately. The performance obtained for the ensemble of random and isolation forests is compared against that obtained by the SVMs -based algorithm.

The results show that the Ensemble of random and isolation Forests (EoF) with graph-based features outperforms the SVMs snf LOF -based algorithm with bag of the system calls as input, for almost all of the attacks in both datasets. This is also confirmed by the ROC curves shown in Figures \ref{fig:roc2022}. The curves, which are obtained by varying the decision threshold in the unsupervised algorithms, highlight the better tradeoff between TPRs and FPRs resulting from the ensemble of forests, since all its curves are located quite close to the upper-left hand corner, while the two baselines produce curves more below with fluctuating trends. By observing the results we can also notice that the detection capability may vary with the type of workload tested; this outcome can be explained by considering that the behavior of some container workloads may hide an underlying attack.

Although the quite high average detection rates, some attacks may yet be able to pass undetected, e.g., true positive rates below 0.7 are resulted against the Backdoor and SQL Injection attacks, and below 0.5 against the Brute Force Login attack.

Finally, there are two attacks (Malicious Script and SQL Injection) where the SVMs-based system outperforms the ensemble of forests in terms of detection rate, nonetheless, the corresponding precision and F1 scores are lower, which is due to the high FPR (Table \ref{tab:fprates}) obtained by the SVMs-based model for dataset Cui-2020.

\section{Conclusion and future works}\label{sec:concl}

In this paper, we presented a new system for online, near real-time intrusion detection for monitoring Docker containers workloads. Our system can be implemented on the hosting machine where the containers are running, and it is based on the graph-based representation of the sequences of system calls generated by the containers. 
We tested the model against different types of anomalous behaviors, and the outcome of our experiments marks the ability to detect the majority of the attacks, despite no labelled attack sample was included in the training set. In addition, our results show that the three stage approach outperforms the two baselines built on one-class SVM and LOF models. Our system is flexible and may be trained on a greater number of expected hosted containers.

Future work will focus on the following: (i) evaluating the scalability of the intrusion detection system by considering an higher number of workloads in the training dataset;  (ii) extending the analysis to other container infrastructures, (iii) investigating the robustness and vulnerabilities of this system against adversarial attacks or other malicious behavior targeting the intrusion detection model; (iv) defining a distributed strategy for training the model and monitoring the containers that may migrate from one hosting machine to another.

\section{Acknowledgements}\label{sec:ack}
This research is partially funded by the EU H2020 ARCADIAN-IoT (Grant ID. 101020259) and partly by the H2020 CONCORDIA (Grant ID. 830927).

\bibliographystyle{IEEEtran}
\bibliography{docbib}

\end{document}